\documentclass{article}
\pdfoutput=1
\usepackage{geometry}
\usepackage[utf8]{inputenc}
\usepackage{graphicx}
\usepackage{lipsum}
\usepackage{float}
\usepackage{xcolor}
\usepackage{cite}
\usepackage{array}   
\usepackage{diagbox}  
\newcolumntype{P}[1]{>{\centering\arraybackslash}p{#1}}

\title{The Adoption of Image-Driven Machine Learning for Microstructure Characterization and Materials Design: A Perspective \\ \vspace{0.2in} \small{This pre-print is currently undergoing peer review for publication in JOM.}}

\author{Arun Baskaran $^1, ^6$ \and
        Elizabeth J. Kautz $^2$ \and
        Aritra Chowdhary $^3$ \and
        Wufei Ma $^4$ \and
        Bulent Yener $^5$ \and
        Daniel J. Lewis $^1$ \\
}

\date{%
    $^1$Department of Materials Sci. \& Engg., Rensselaer Polytechnic Institute, NY, USA\\%
    $^2$Pacific Northwest National Laboratory, WA, USA\\%
    $^3$Artificial Intelligence, GE Research, NY, USA \\
    $^4$Department of Computer Science, Purdue University, IN, USA \\
    $^5$Department of Computer Science, Rensselaer Polytechnic Institute, NY, USA\\
    $^6$Currently at the Center for Nanoscale Materials, Argonne National Laboratory, IL, USA
}

\begin{document}

\maketitle

\begin{abstract}

The recent surge in the adoption of machine learning techniques for materials design, discovery, and characterization has resulted in an increased interest and application of Image Driven Machine Learning (IDML) approaches. 
In this work, we review the application of IDML to the field of materials characterization. 
A hierarchy of six action steps is defined which compartmentalizes a problem statement into well-defined modules.
The studies reviewed in this work are analyzed through the decisions adopted by them at each of these steps. 
Such a review permits a granular assessment of the field, for example the impact of IDML on materials characterization at the nanoscale, the number of images in a typical dataset required to train a semantic segmentation model on electron microscopy images, the prevalence of transfer learning in the domain, etc. 
Finally, we discuss the importance of interpretability and explainability, and provide an overview of two emerging techniques in the field: semantic segmentation and generative adversarial networks.

\end{abstract}

\section{Introduction}
\label{intro}
\color{black}

\textit{Microstructure} is a broad term that encompasses the description of a material's structure at spatial length scales ranging from millimeters to nanometers.
A material's property and performance is strongly influenced by its microstructure, which is often studied by capturing and analyzing images using various microscopy methods. 
Such micrographs are rich in information about the material's origin and processing history and its chemical make-up. 
Hence, microstructure visualization, analysis, and interpretation has become ubiquitous in materials science research. 
Significant advances in both experimental and computational methods have enabled advances in how we visualize and describe material microstructures. 
For example, three dimensional microstructures can be visualized with techniques such as serial sectioning and atom probe tomography \cite{Alkemper2001, DevarajReview}, and in situ methods have been developed for direct visualization of atomic-scale phenomena, including oxide growth mechanisms and element redistribution \cite{LangliSciAdv, KautzScripta}. 
Computational methods have been at the forefront of many materials research studies due to the need for accelerated materials discovery, design, and development emphasized in the Materials Genome Initiative \cite{McDowell2016}. 
Methods including ab initio calculations, density functional theory (DFT), n-point statistics, and machine learning have recently gained more widespread use, enabled by modern (often, open source) algorithms and high-performance computing. 
More recently, the application of artificial intelligence (AI) has grown significantly in engineering and science domains, particularly in the materials science area. 
This boom in AI has been referred to as the fourth paradigm of science \cite{Agrawal2016} and the fourth industrial revolution \cite{Schwab2015}, and has enormous potential in altering how materials scientists discover new material systems, predict properties, and even interpret and analyze micrographs.

The topic area of machine learning in materials science, or material informatics, is broad. 
Reviews of the overall field can be found in \cite{Rickman2019,Ramprasad2017}. 
Within this broad field, a specific area of importance is the characterization and interpretation of image data using machine learning methods, referred to here as image driven machine learning (IDML). 
Image data is ubiquitous in materials characterization efforts, thus, there is a strong need to develop models and approaches that can accurately, effectively, and reliably link image data to other parameters of interest (i.e. processing parameters, properties, etc.). 
In the specific area of microstructure image analysis, machine learning is proving highly valuable for characterizing diverse microstructure data sets paramount to the materials design and discovery process \cite{DeCost2015,Chowdhury2016}. 
Yet, there is still much to be learned about how to apply machine learning methods to the important tasks of microstructure recognition, characterization, development of processing-microstructure-property relationships, and microstructure design.  
The motivation of the current review is to provide a perspective on the different aspects involved in the application of IDML techniques for material characterization, by looking at the existing literature.
The text is likely to provide a high level overview of the field to material scientists who are interested in the field of applied machine learning, as well as enable the experienced practitioners to obtain a summary of the field or an insight into an allied field.  \color{black}

In this article, we begin by outlining the current state-of-the-art of IDML in materials science. 
The scope of this review includes image datasets obtained from camera (imaging sample surface at the resolution of naked eye), optical microscopy, electron microscopy, spectroscopy, diffraction patterns, and from simulations of material structure at different length scales. 
Following this, we define a canonical hierarchy of procedural stages which is used to modularize the IDML studies reviewed in this work. We briefly discuss the importance of interpretability and explainability, towards a widespread adoption of IDML in materials science. 
Finally, we briefly discuss two emerging techniques that have gained importance in the last couple of years: generative adversarial network and semantic segmentation.

\section{Overview of the field}
\label{overview}
Although it is common for materials scientists to relate microstructure and properties, the availability of microstructural information is a relatively recent discovery compared to the length of time materials technology has been known.
C. S. Smith\cite{CSSmith} has written extensively on the subject of the discovery of microstructure in materials and cites what is considered to be the first observations of metallic crystallanity through the use of chemical etchants in meteorites by Thompson in 1804 and in Damascus steel by Breant in 1821.
The need to quantify the description of these structures was immediately apparent and we refer the reader to C. S. Smith's writings on the history of technology and structure of metals.
A significant contribution of Smith's was in the description of polycrystalline structures and an early application of a computational image processing technique to a microscopy image was the use of the intercept method \cite{Abrams1971} to calculate the average grain size of a polycrystal.
Since then, the field has experienced the use of progressively advanced techniques to extract complex insights from microscopy images.
While the next section will review journal articles in this field on the basis of their technical aspects, this section will focus on the materials science problems that have been addressed so far using IDML.

\begin{figure}[htbp]
    \centering
    \includegraphics[width = 0.45\textwidth]{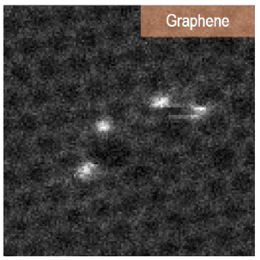}
    \includegraphics[width = 0.45\textwidth]{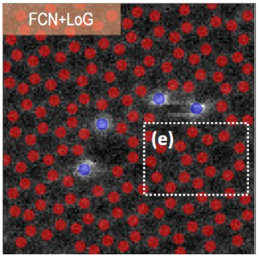}
    \includegraphics[width = \textwidth]{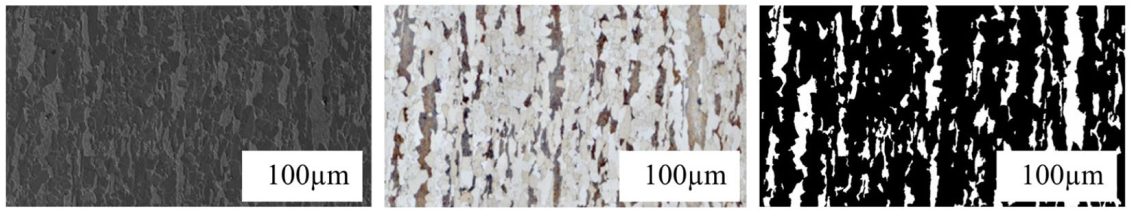}
    \caption{IDML has influenced material analysis across a range of magnifications. (Top) On the smaller end of the spatial length scale, studies have targeted microscopy images with a field of view spanning between 10-500 nm. Figure reproduced with permission from \cite{Ziatdinov2017}; Copyright American Chemical Society (2017). (Bottom) On the larger end of the spatial length scale, studies have targeted microscopy images with a field of view of 10 $\mu m$ or more. Figure reproduced with permission from \cite{Azimi2018}; Creative Commons CC BY.}
    \label{fig:magnifications}
\end{figure}

%
In the last decade, IDML has augmented the human expert's understanding of material characterization at a range of length scales spanning multiple orders of magnitudes.
At the lower magnification levels, IDML techniques have been applied to image datasets with a field of view in the range of 10-100 mm.
Examples of such studies include porous microstructure reconstruction using generative methods \cite{Zhang2021}, in-situ defect detection in melt pools for additive manufacturing \cite{Gobert2018,Scime2019}, and segmentation of images generated from x-ray computed tomography and serial sectioning \cite{Stan2020}.
At comparatively higher magnification levels,  IDML has been used to analyze optical and electron microscopy image datasets with a field of view in the range of 10-500 $\mu m$.
A few examples of such studies include high level classification of images into an appropriate microstructure class(\cite{Chowdhury2016,DeCost2015}), semantic segmentation of images generated from x-ray computed tomography (\cite{Strohmann2019,Evsevleev2020}), semantic segmentation of images generated from Scanning Electron Microscopy (\cite{Tsopanidis2020,Azimi2018}), quantitative metallographic analysis of microscopy images (\cite{Campbell2018,Agbozo2019}).
At the high end of magnification, IDML has been used to analyze images with a field of view in the range of 10-500 nm.
Examples of such studies include high-level classification of chirality of carbon nanotubes from images generated by High-Resolution Transmission Electron Microscopy \cite{Forster2020}, semantic segmentation of crystallographic defects from images generated by Scanning Transmission Electron Microscopy (\cite{Ziatdinov2017,Roberts2019}, nanoparticle segmentation from images generated by liquid-phase Transmission Electron Microscopy \cite{Yao2020}, and analyzing the grain boundary character  \cite{Wei2019_APTML}.
A visualization of characteristic images spanning the length scales discussed above has been shown in Figure \ref{fig:magnifications}.

A different perspective about the field can be obtained by extracting the domain specific problem statements from the studies. 
The most extensive application of IDML in the domain has been towards phase segmentation and quantitative analysis of microstructures.
The objectives of a segmentation study include the isolation of specific phases \cite{Azimi2018,Evsevleev2020} and the extraction of quantitative information, such as volume fraction, about a specific phase \cite{Agbozo2019, Chan2020}.
In \cite{Madireddy2019}, a supervised approach for binary edge detection allowed for segmentation of phases with different morphologies than were present in the training dataset. \color{black}
Another common application of IDML in this field is the assigninment to an image a feature label that best describes the image, from a pre-determined candidate set of features exhibited by the material.
The label can indicate a morphology present in a microstructure image (\cite{Baskaran2020,Chowdhury2016,DeCost2015}), the chirality exhibited by the sample \cite{Forster2020}, processing history of a given microstructure \cite{Ma2020_1}, or crystal structure identification from a Transmission Electron Microscopy image \cite{Aguiar2019}.
Assigning a class label can providing the human expert with a high-level overview of a microstructure or be an integral part of a multi-stage machine learning pipeline.
IDML has also been used for the purpose of defect detection at different length scales and quality control.
Examples of such studies include in-situ defect detection in additive manufacturing \cite{Scime2019,Gobert2018}, fracture surface analysis from electron microscopy images \cite{Tsopanidis2020}, and segmentation of crystallographic defects from electron microscopy images \cite{Ziatdinov2017,Roberts2019}.

\subsection{Related Work}

Machine learning, and more specifically the IDML framework, is considered here to be a subset of the larger domain of Materials Informatics. 
It is important to note that the terms data-driven, machine learning, and materials informatics are typically used interchangeably. 
Materials informatics, however, is a large field that encompasses machine learning and statistical models for the quantitative description of microstructures. 
Several prior works have implemented statistical characterization and reconstruction of microstructures \cite{Bostanabad2018,Paulson2017,Torquato1982,Kalidindi2011,Chen2019}. 
Another important component of microstructure quantification is the uncertainty quantification, specifically the processing-induced variability in features and its subsequent effect on the material properties. Examples of studies on this topic include \cite{Acar2017, Huang2021}. \color{black}
While these studies are incredibly valuable and insightful, the application of ML methods for microstructure quantification offer some unique advantages over these statistical techniques, including the ability to incorporate large amounts of data about the material microstructure and analysis of never-before seen data.

Machine learning is now being increasingly used in the materials science domain for a wide range of applications, beyond analysis/classification of images. Although IDML is the focus of this article, it is important to mention that image classification and analysis is merely a subset of the broader field of machine learning in materials science. Hence, we briefly summarize related work on machine learning methods (e.g., neural networks, Support Vector Machines, k-means clustering, random forests, generative networks, and more) applied to several diverse challenges in molecular and materials science fields. In particular, active research areas for ML in materials science include (but are not limited to): accelerated materials design and property prediction \cite{Nikolaev2016, Ye2018, Oses2018, Draxl2018, Plata2017, Kalinin2016, Kautz2019TC},  process optimization \cite{Liu2020_acta, Fang2009}, discovery of structure-property relationships \cite{Yang2018, Seko2018}, construction of potential energy surfaces for molecular dynamics simulations \cite{Ballard2017, Chmiela2018, Jose2012}, prediction of atomic scale properties \cite{Bartok2017}, text mining for knowledge extraction\cite{Kim2017}, microstructure and materials characterization \cite{Ling2017, DeCost2015, Kautz2020_IDML, Baskaran2020, Chowdhury2016, Park2017, Ziatdinov2017}, and generation of synthetic microstructure images \cite{Ma2020_1}. Such applications span multiple length scales and a variety of material systems (metals and alloys, oxides, polymers) \cite{Liu2017}.

The merger of machine learning with materials science is a relatively new and growing field that contributes to the evolution from traditional methodologies, where experimental characterization techniques were used to understand processing-structure-property relationships, to one that is data-driven. This paradigm shift can, in many cases, help accelerate materials science research through a more autonomous, objective, and reliable design and characterization process, which is impacted less by researcher bias and chance discovery \cite{Butler2018}.

The ability to analyze smaller data sets is important in materials science where a limited number of micrographs are usually available for any single study. Recent advances have led to developments that allow human-level performance in one-shot, or few-shot learning problems \cite{Butler2018, Lake2015}. Typically, data analysis via deep learning methods require large amounts of data (such as the large image data available through the ImageNet database \cite{Krizhevsky2012,Deng2009}). Although in many materials science studies, researchers are limited to a few data points, an advantage of many micrographs is that several microstructural features of interest are available for interest in a single image (depending on magnification). This one-shot or few-shot learning concept has significant implications for future materials science studies, for example characterization of neutron irradiation or corrosion effects, where limited material is available for analysis due to long lead-time experimentation. In other cases, there exists data from previous studies that may be very limited, or not well understood, for which advanced data analysis methods could be applied, as was done in prior work \cite{Kautz2019TC, Mace2018}.

This relatively recent surge in research efforts at the intersection between materials science and machine learning domains have been enabled by the ever decreasing cost of computing resources. In addition, several frameworks such as Keras\cite{Chollet2015}, PyTorch \cite{Paszke2019},\color{black} and TensorFlow \cite{Abadi2015} have helped materials scientists (amongst other domain scientists such as nuclear engineers, chemists, physicists, etc.) and data scientists readily apply sophisticated machine learning algorithms with relatively low/minimal developmental efforts. The low cost associated with application of machine learning algorithms to various domain-specific research challenges, and the aforementioned benefits enable studies such as those discussed in this article.

\section{Component-based breakdown of IDML research efforts}
\label{IDML_current_work}
\begin{figure}[htbp]
    \centering
    \includegraphics[width=0.95\linewidth]{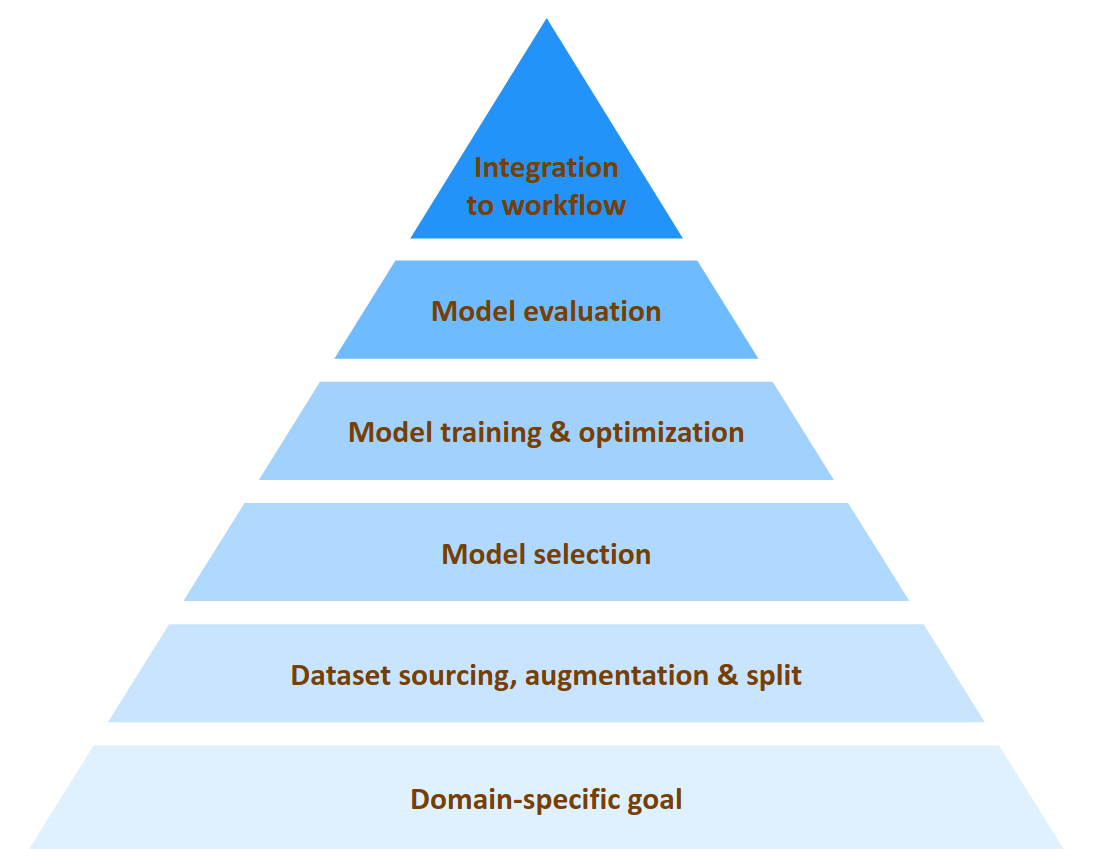}
    \caption{The canonical series of procedural stages used to review the studies in the field of IDML for material characterization. Each stage can be considered as being independent of the others, but crucial to the final performance of the model or algorithm that is being implemented.}
    \label{fig:hierarchy}
\end{figure}

This section provides a modularized approach to discuss the published literature in the field of IDML for materials characterization.  
A series of six canonical action steps is defined in Figure \ref{fig:hierarchy}.
The review is divided into six components, one for each of the steps outlined in the figure.
The granularity of the modularization is chosen so as to keep the individual modules comparable to application of ML in allied fields.
%

%
Each action step in the series is approached sequentially, from the bottom-up, in the logical order required to successfully implement a ML technique to the domain.
The first step in the series is the definition of a domain-specific goal, i.e., the problem specific to materials science that is being answered by the adoption of IDML. 
An appropriate problem definition is important to assess the degree of success of the implementation. 
A review of the field from the perspective of this step is likely to help a beginner to the field with identifying the different components that together make up a rigorous problem definition. well as an experienced practitioner.  
For an experienced practitioner, this component of the review provides a summary of the domain-specific problems that have been approached through IDML methods.

The second step in the series is dataset acquisition. Data collection, augmentation, and pre-processing are important steps for a project implementing both computer vision and machine learning techniques. In addition, the allocation of data into training, validation, and testing have also been discussed where appropriate.

Following this, an overview is provided of the classification models and image processing algorithms used among the published literature in the field. Specific focus is provided to the rationale behind model selection that is appropriate for a particular domain-specific goal and dataset. Specific model-based tuning to address domain-specific requirements are highlighted in this subsection.

The fourth component of the review is the training and optimization methods employed by the published studies in the domain. Here as well, the specific modifications to the training process to address domain-specific constraints are highlighted.

The fifth component of the review is the evaluation metrics adopting in the field to assess the performance of the IDML techniques. Evaluation metrics are an assessment of the extent to which the technique addresses a given materials characterization analysis, and hence a greater focus is given to this section. This section is aimed at giving the reader an overview of the benchmark results currently in the field, as well as providing an introduction to a set of metrics that are commonly used in the field.

The final component of the review pertains to a discussion of ways by which research efforts have integrated IDML into their characterization workflows.
\label{IDML_components}

\subsection{Domain-specific goals}
\label{functional_goals}

Table \ref{tab:functional_goals} outlines the functional goals outlined in the 30 papers reviewed in this study.

\begin{table}
    \centering
    \caption{Domain-specific goals adopted by a subset of studies that have been reviewed in this work. }
    \begin{tabular}{|P{10em}|P{32em}|}
        \hline
        Paper & Functional goals \\
        \hline
        DeCost2015 \cite{DeCost2015} & Classify microstructure images into one of seven image classes, and develop a bag of visual features \\
        \hline
        Chowdhury2016 \cite{Chowdhury2016} & Binary classification of microstructural features (dendrites) and feature orientations \\
        \hline
        Zhang2021 \cite{Zhang2021} & Generate isotropic and anisotropic synthetic 3D porous structures using 2D slices as input  \\
        \hline
        Gobert2018 \cite{Gobert2018} & In situ defect detection detection using supervised machine learning, for powder bed fusion (PBF) additive manufacturing \\
        \hline
        Scime2019 \cite{Scime2019} & In-situ defect detection in additive manufacturing melt pools \\
        \hline
        Stan2020 \cite{Stan2020} & Semantic segmentation of computed tomography and serial sectioning images \\
        \hline
        Strohmann2019 \cite{Strohmann2019} & Semantic segmentation of 3D microstructure of Al-Si  \\
        \hline
        Evsevleev2020 \cite{Evsevleev2020} & Deep-learning based semantic segmentation of individual phases from synchrotron x-ray computed tomography images \\
        \hline
        Tsopanidis2019 \cite{Tsopanidis2020} & Semantic segmentation of fracture images of MgAl2O4, using a deep-cnn \\
        \hline
        Azimi2018 \cite{Azimi2018} & Pixel-wise segmentation of steel microstructure datasets \\
        \hline
        Campbell2018 \cite{Campbell2018} & Automated extraction of quantitative data from material microstructures using advanced image processing technique \\
        \hline
        Agbozo2019 \cite{Agbozo2019} & Quantitative metallographic analysis through object segmentation of SEM images \\
        \hline
        Forster2020 \cite{Forster2020} & Classification of HRTEM images of Carbon Nanotubes into its appropriate chirality  \\
        \hline
        Ziatdinov2017 \cite{Ziatdinov2017} & Semantic segmentation of defects and subsequent defect structure identification from atomic scale STEM images \\
        \hline
        Roberts2019 \cite{Roberts2019} & Semantic segmentation of crystallographic defects from electron micrographs\\
        \hline
        Yao2020 \cite{Yao2020} & Nanoparticle segmentation from liquid-phase TEM images by U-Net \\
        \hline
        Chan2020 \cite{Chan2020} & Automated quantitative analysis of microstructure using unsupervised algorithms \\
        \hline
        Baskaran2020 \cite{Baskaran2020} & Contextual segmentation of morphological features in titanium alloys using a two-stage machine learning pipeline  \\
        \hline
        Aguiar2019 \cite{Aguiar2019} & Classification of TEM data, atomic resolution images and diffraction data into crystal structures at family level and genera level \\
        \hline
        Zhu2017 \cite{Zhu2017} & Particle recognition from cryo-EM datasets\\
        \hline
        Chen2020 \cite{Chen2020} &  Instance semantic segmentation of Al alloy metallographic images\\
        \hline
        DeCost2018 \cite{Decost2018} & Semantic segmentation of ultrahigh carbon steel microstructures through deep learning techniques \\
        \hline
        Furat2019 \cite{Furat2019} & Semantic segmentation of computed tomography data of Al-Cu specimens \\
        \hline
        Vuola2019 \cite{Vuola2019} & Nuclei segmentation from microscopy images for biomedical imaging using an ensemble model\\
        \hline
        Hwang2020 \cite{Hwang2020} &  Semantic segmentation of multi-phase composite microstructures \\
        \hline
        Ma2018 \cite{Ma2018} & Semantic segmentation of Al-La microstructure images \\
        \hline
        Decost2017 \cite{DeCost2017} & Classify AM feedstock powder images into their respective material system \\
        \hline
        Yang2019 \cite{Yang2019} & Prediction of the microscale elastic strain field in a 3D voxel-based microstructure \\
        \hline
        Cang2016 \cite{Cang2016} & Microstructure reconstruction from a low-dimensional representation\\
        \hline
        Arganda-Carreras2017 \cite{Arganda-Carreras2017} & Develop a library for integration of machine learning methods available in WEKA into image processing toolkit Fiji for biological and non-biological specimens.\\
        \hline
        Ciresan2012 \cite{Ciresan2012} & Binary membrane segmentation of neuronal structures in stacks of electron microscopy images \\
        \hline
        Wang2013 \cite{Wang2013} & Binary membrane segmentation of neuronal structures in stacks of electron microscopy images\\
        \hline
        Haan 2019 \cite{Haan2019} & Super-resolving low-resolution SEM images through a generative adversarial network \\
        \hline
        Kaufmann2020 \cite{Kaufmann2020} & Crystal structure identification from EBSD diffraction patterns using deep learning \\
        \hline
        Madsen2018 \cite{Madsen2018} & Semantic segmentation of HRTEM images for local structure recognition \\
        \hline
    \end{tabular}
    \label{tab:functional_goals}
\end{table}

\begin{figure}[htbp]
    \centering
    \includegraphics[width = \textwidth, height = 1.25in]{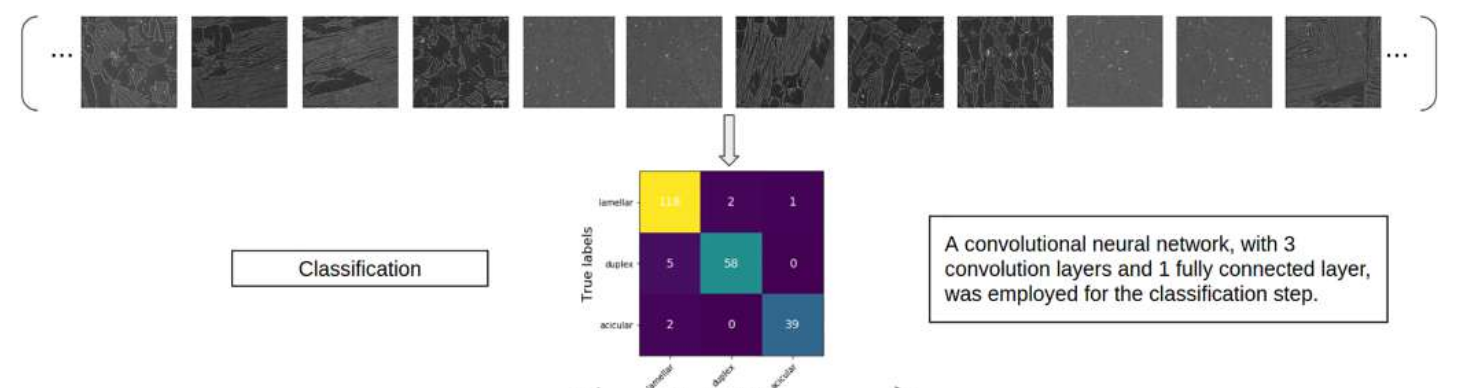}
    \includegraphics[width = 0.6\textwidth, height=2.5in]{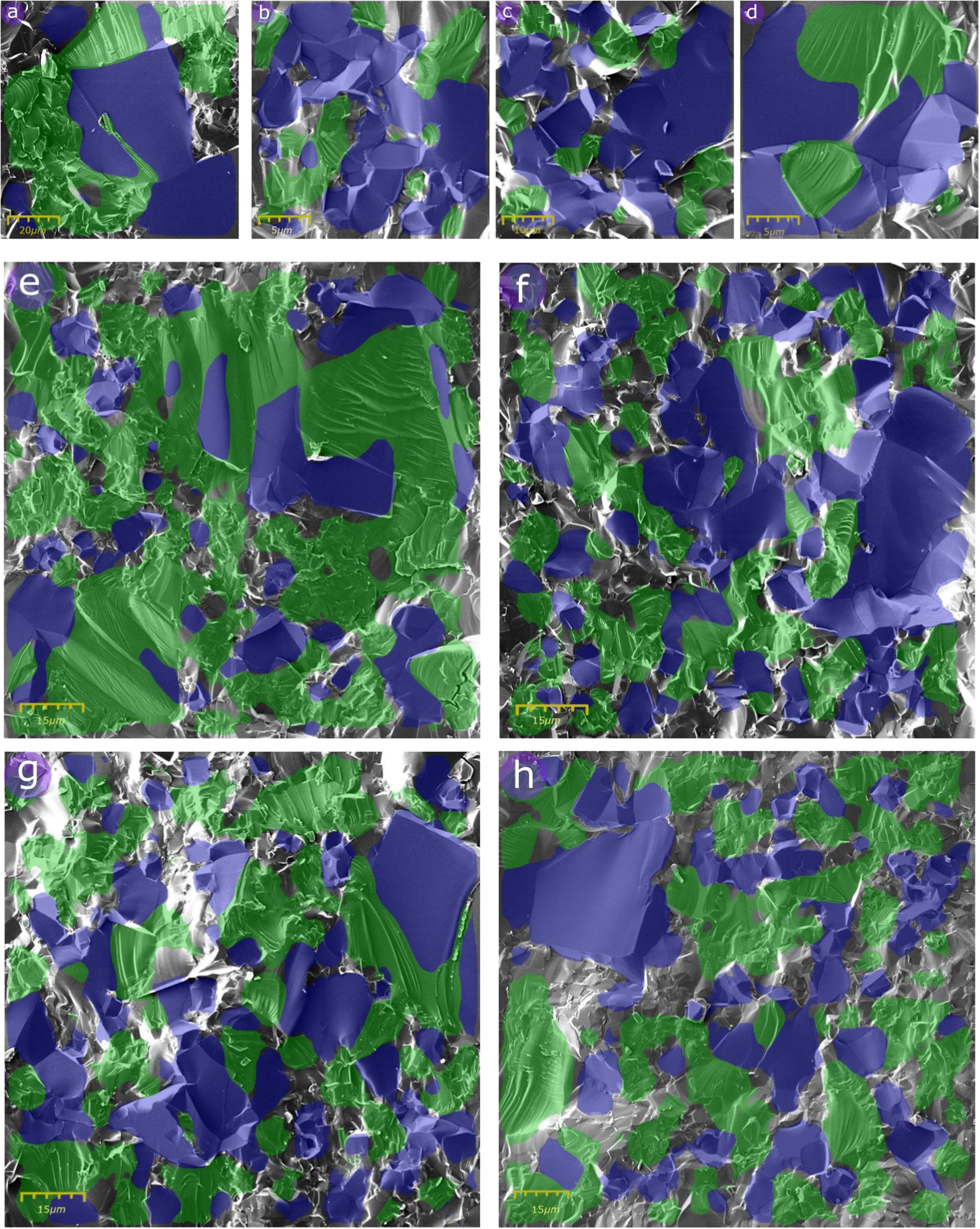}
    \caption{Distinction in the types of classification that can be performed, based on the number of pixels per classification label. (Top) Coarse classification, in which a label is assigned to each image in the dataset. Figure reproduced with permission from \cite{Baskaran2020}; Copyright Elsevier (2020). The example image is extracted from \cite{Baskaran2020}. (Bottom) Dense classification (or pixel-based classification), in which a label is assigned to each pixel in the image. Figure reproduced with permission from \cite{Tsopanidis2020}; Copyright Elsevier (2020).}
    \label{fig:image_scale}
\end{figure}

One of the main goals of material scientists in relation to the use of computer vision techniques has been towards microstructure classification. The research efforts in this area can be divided into two broad categories, coarse classification and dense classification, depending on the number of pixels per classification label. The first category includes classification models that assign one label per image in the dataset. The label that is chosen is the one that best represents the microstructure in the image, such as ``dendritic'', ``lamellar'', etc. Examples of previous studies that perform such a classification include \cite{DeCost2015}, \cite{Chowdhury2016}, \cite{Azimi2018}, \cite{Baskaran2020}, and \cite{Tsutsui2019}. Dense classification models include typically assign one label per pixel in a given image. In this case, the label describes the phase of the microstructure at the spatial location represented by the pixel. Examples of research articles that perform this classification include \cite{Stan2020,Strohmann2019,Evsevleev2020,Tsopanidis2020,Campbell2018,Agbozo2019,Chan2020,
Zhu2017,Chen2020,Decost2018,Furat2019,Vuola2019,Hwang2020,Ma2018,Wang2019}. 
A fast emerging technique in this category is semantic segmentation, which can be summarized as pixel classification performed in the context of the whole image. It is noted that there also exists models that perform classification at a scale that is in between the two extremes stated above, such as the prediction of bounding boxes using a Region Proposal Network (RPN) in \cite{Chen2020}. 
Figure \ref{fig:image_scale} shows characteristic examples of application of the different types of classification discussed above to characterization analysis.

These styles of classification have been used by the domain experts towards two broad objectives, object/class detection and extraction of quantitative information from characterization results. The object (for example, precipitate) or class (for example, dendritic) detection is typically performed by either coarse-grained classification or semantic segmentation. Another use for object detection has been the detection of defects and abnormalities. The goal for such a study would be binary classification into ``defect'' and ``no defect''. The classification result is typically integrated with the feedback for process-correction or quality-control. Examples of research articles with this goal include \cite{Gobert2018,Scime2019,Roberts2019,DeCost2017,Burger2014,Lorenzoni2020,Liu2019}. Among the research articles cited above as examples of pixel-level classification, one of the goals is to aggregate the classification into developing a quantitative finger-print of the microstructure, such as area fraction of a particular phase or the number of precipitates in the field of view represented by the image. Both object detection and quantitative analysis of the microstructure potentially aid the study of a process-structure relationship for the material.

Among the studies that have performed quantitative microstructure analysis, a subset of them have developed a lower-dimensional representation for microstructures. Techniques like Principal Component Analysis are used to identify the most discriminative features for the dataset. Such a representation could be an end-goal in itself, like in \cite{Cang2016,Han2020,Lubbers2017,Niezgoda2013,Tsutsui2019,Xu2014,Xu2015}, or could be developed in order to improve the classification performance, like in \cite{Chowdhury2016,Ma2020_1}.

\subsection{Datasets}
\label{Dataset_collection}


The dataset characteristics, such as its size and distribution, have a strong influence on the overall performance of the model and/or algorithm.
Most applications of IDML in the materials science domain (not including the studies involving microstructure reconstruction) were performed on real data, i.e. data generated from real material samples.
A few exceptions to this can be found in studies such as \cite{Yang2019,Ma2020_1,Forster2020,Aguiar2019,Ziatdinov2017}, where simulated data either comprised the entire training dataset or supplemented the data obtained from a real sample.
The real data were generated from a wide variety of sources, which have been briefly summarized in Section \ref{overview}.
One can observe a significant variation in dataset sizes among the published literature in this field.
There is no rigid rule to determine the appropriate size of the training dataset. 
It depends on factors such as the problem statement, variation in the dataset, choice of model, required level of performance, etc. 
Naturally, coarse classification and dense classification studies require different number of images for training.
Since the former has a greater number of pixels per classification label than the latter, it requires a comparatively larger dataset.
Even among studies that have employed a coarse classification model, a significant variation in dataset size can be observed.
For example, while \cite{DeCost2015} uses 105 images to train a SVM for classification into 7 classes, \cite{Chowdhury2016} employs 528 images for a binary classification task using different models, and \cite{Baskaran2020} uses 1225 images to train a CNN for 3-class classification.
To the best of our knowledge, the largest dataset used in the domain of materials characterization has 1.3e+6 images and is reported in \cite{Forster2020}.
Similarly, a variation in dataset sizes can also be observed among studies that have used dense classification models.
For example, while the reported datasets in \cite{Agbozo2019,Chen2020,Decost2018} are 50, 100, and 24 respectively, the reported datasets in \cite{Azimi2018,Ziatdinov2017} are 5093 and 2000, respectively.

\begin{figure}[htbp]
    \centering
    \includegraphics[width=0.45\textwidth]{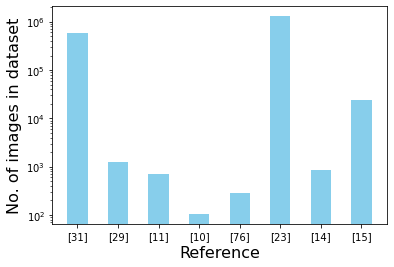}
    \includegraphics[width=0.45\textwidth]{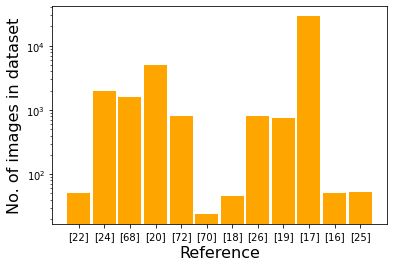}
    \caption{A representation of the dataset sizes used by a subset of studies reviewed in this work. Since coarse classification and pixel-level classification have dataset requirements spanning different orders of magnitude, they are represented in separate charts}
    \label{fig:databars}
\end{figure}

In addition to the size of the overall dataset, it is also important to appropriately split the dataset for the purposes of training, validation, and testing. A good evaluation of the performance of a trained model is its performance on data that was not used for training.
A general trend observed in the studies reviewed in this work is that the data split depends on the size of the overall dataset.
When the dataset size is comparatively large, such as in \cite{Azimi2018,Aguiar2019}, the authors have allocated data for a held-out dataset to evaluate the performance of their trained models.
However when the dataset is comparatively small, like in \cite{Decost2018,Chowdhury2016}, the performance of the trained model is evaluated using a cross-validation technique.
The evaluation metrics are discussed in detail in the next subsection.
A common challenge faced by the data acquisition methods in the field is that it is time-consuming and requires expertise to assemble a large, high-quality, image dataset that can effectively train a machine learning model.
Data augmentation has been widely used to overcome this challenge.
Among studies reviewed in this work, some of the common augmentation techniques include cropping \cite{Baskaran2020,Tsopanidis2020,Azimi2018}, rotation \cite{Strohmann2019,Agbozo2019,Zhu2017,Evsevleev2020}, and translation \cite{Strohmann2019}
As mentioned previosusly, simulated data have also been used to supplement the data from real imaging sources.
Figure \ref{fig:databars} lists the dataset characteristics observed in the studies reviewed in the current work.




%

\subsection{Model selection and training}
The selection of the appropriate model for a given application depends on factors such as the the problem statement, distribution and size of the dataset, the computational resources available, etc.
The field of IDML for material characterization has adopted a wide variety of models, ranging from shallow to deep learning.\color{black}
Reflecting the current state of the field, the models can be categorized into two distinct categories: non-neural network and neural network.
The former belongs to a shallow learning paradigm, whereas the latter belongs to a deep learning paradigm.
Shallow learning refers to a learning paradigm in which the model parameters (for example, the weights of support vector machine model) are learnt directly from the input features.
In other words, the input features are connected directly to the output via the model.
In contrast, deep learning refers to a learning paradigm in which the input features are sequentially fed to one or more intermediate layers(also known as hidden layers) before an output is obtained.
In this paradigm, only the first intermediate layer learns directly from the input features.
The hidden layers learn an alternate, often lower-dimensional, representation of the dataset. 
Owing to model simplicity and the fewer number of trainable parameters, shallow learning models are easy to train and can provide reasonable performance with a small training dataset.
However, the training process is strictly empirical and it is less likely to learn hidden patterns in the dataset. 
On the contrary, a deep learning method is computationally expensive to train and requires a relatively larger dataset for training compared to shallow learning models.
However, the trained hidden layers can provide an insight into the underlying patterns in the dataset, and can be used to explain the predictions of the model.

Coarse-grained classification tasks that generate one class label per image have been demonstrated with shallow learning algorithms such as Support Vector Machine \cite{Chowdhury2016,Scime2019,DeCost2017,Gobert2018} and Random Forest Classifier \cite{Chowdhury2016,Ma2020_1}.
Reflecting the growing popularity of the convolutional neural network (CNN) in related fields, there has been a push in the application of CNNs for analyzing microstructure images.
This trend can be attributed to two main factors: improved performance for multi-label classification and its application for semantic segmentation.
{It is important to note that CNNs and deep CNNs (DCNNs) are often selected, as opposed to multi-layered perceptron networks (MLPs) because the image data input is conducive to CNNs, which requires a spatially-dependent data format \cite{Goodfellow2016}. CNNs and DCNNs take input image data and convolve intermediate image data with learned kernels in several successive layers, allowing for the network to learn highly nonlinear features.}

\begin{figure}[htbp]
    \centering
    \includegraphics[width=0.48\textwidth, height=1.0in]{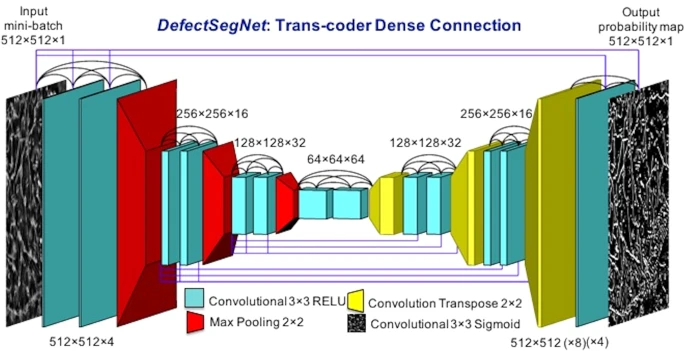}
    \hspace{0.02\textwidth}
    \includegraphics[width=0.48\textwidth, height=1.0in]{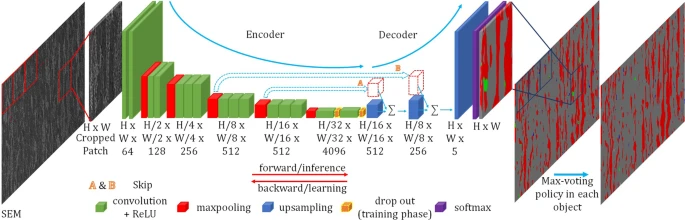}\\
    \vspace{0.2in}
    \includegraphics[width=0.8\textwidth, height=1.75in]{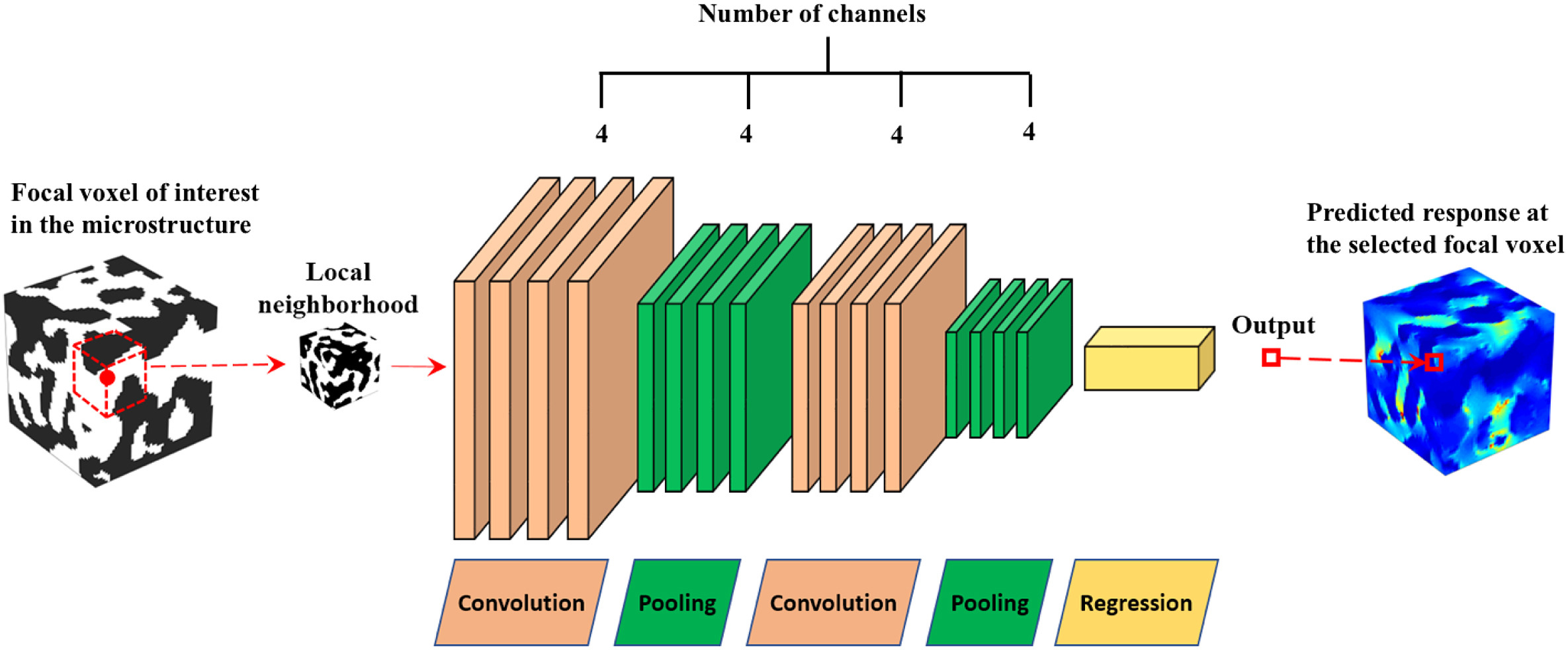}
    \hspace{0.02\textwidth}
    \vspace{0.05in}
    \includegraphics[width=0.8\textwidth,height=2.0in]{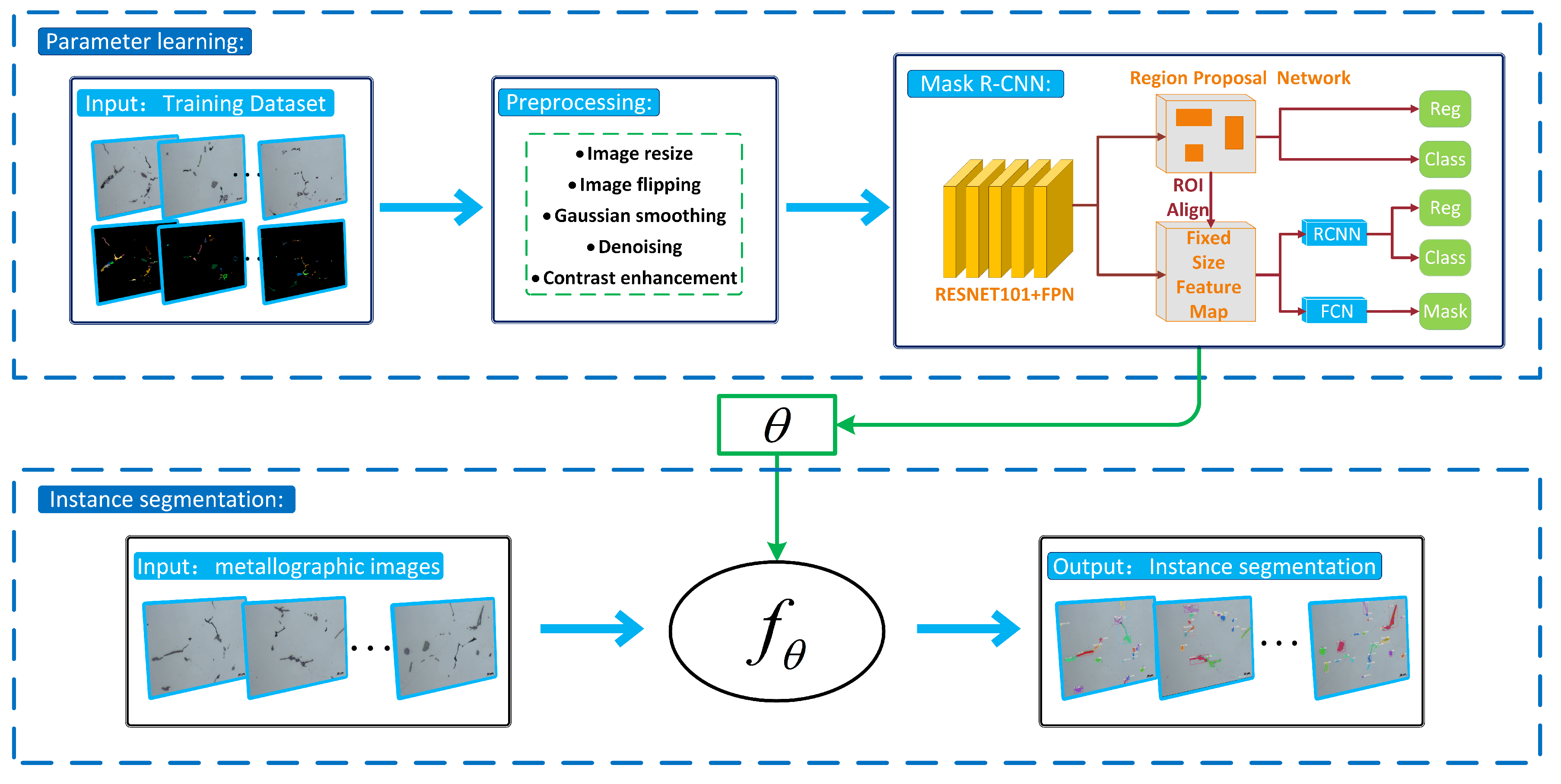}
    \caption{A selection of models employed by studies reviewed in this work. From top left, the figure were reproduced with permission from \cite{Roberts2019}; Creative Commons CC BY, \cite{Azimi2018}; Creative Commons CC BY, \cite{Yang2019}; Copyright Elsevier (2019), and \cite{Chen2020}; Creative Common CC BY.}
    \label{fig:models}
\end{figure}

The design of a typical CNN is such that it learns about inherent patterns, through trainable filters, at different spatial resolutions compared to the input image.
Libraries, such as Keras \cite{Chollet2015}, enable a visualization of these patterns \cite{Baskaran2020} and thus can provide insight into the rationale behind the model's predictions.
A few examples that perform coarse classification using CNNs include studies detailed in \cite{Aguiar2019,Baskaran2020,Forster2020}.
The rising popularity of deep learning techniques for semantic segmentation applications in allied fields\cite{Garcia-Garcia2017} is reflected in an increased adoption of neural networks for pixel-based classification of material images.
{It should be noted a typical neural network architecture for coarse classification is different from a typical neural network architecture for pixel-level classification.
For the former, the neural network (CNN) progressively downsamples the input image and provides as output a vector of class probabilities.
The part of the model that is responsible for downsampling an input image is typically known as an encoder.}
For a pixel-level classification, the encoder is connected to a decoder which upsamples the lower resolution image to provide an output of the original resolution.
Examples of studies that have used an encoder-decoder structure include \cite{Agbozo2019,Azimi2018,Chen2020,Decost2018,Furat2019,Vuola2019}. \color{black}
%
The emerging field of semantic segmentation is discussed in a later section.
Notable advancements in neural network architecture that have been adopted by the material science community include the use of Mask R-CNN for instance segmentation \cite{Chen2020}, use of an architecture with atrous convolution layers \cite{Hwang2020,Ma2018}, use of skip connections \cite{Roberts2019}, etc.
A few examples of model architectures used in the field have been highlighted in Figure \ref{fig:models}. 
The availability of trainable filters have also helped in augmenting domain-specific knowledge into the training process \cite{Chidester2019,Marcos2016}.

Depending on the availability of data, two different training modes can be implemented. 
In the first mode, the model is trained from a random initial state, i.e., the parameters of the model such as the filter weights are initialized to a specified random distribution. 
During the training period, these parameters converge to values that are optimal for the dataset. 
he number of training samples for this mode scales with the number of parameters, and hence it may not be feasible to implement complex deep learning models from scratch to data-deficient applications. 
Previously implemented works which have performed training from scratch include \cite{Baskaran2020,Ziatdinov2017,Zhu2017,Yang2019}.
The second training mode is referred to as transfer learning, and takes advantage of models that have been pre-trained on large datasets. 
In this mode, the model parameters are initialized to the values that were obtained at the end of the pre-training. 
Subsequently, a part or whole of the model is fine-tuned with the dataset in hand. 
It is not common to directly use the pre-trained model parameters without fine-tuning, unless the current dataset and the dataset for pre-training are similar in nature. 
A major advantage of the transfer learning is that complex deep learning models can be implemented using comparatively fewer training data.
Previously implemented works which have performed transfer learning include \cite{Agbozo2019,Decost2018,Vuola2019,Tsopanidis2020}.
An emerging sub-field within transfer learning is few-shot learning \cite{Akers2021}, in which a pre-trained model is fine-tuned using a very small training dataset (often ~ 5 images). 
Few shot learning in particular, and transfer learning in general, is especially beneficial in cases where images dissimilar to the training dataset are encountered.
In addition, transfer learning also reduces the demand for computational resources imposed by large models, and hence is likely to help in implementing machine learning at the edge (for example, in individual computers connected to microscopes). 
The reader can refer to open source libraries such as \cite{seg_models} which have made it easy to load pre-trained states of commonly used models.

In some cases, the bottleneck may be in the generation of ground truth data for supervised training.
One way to overcome this is to explore the possibility of unsupervised training, i.e., training without the presence of ground truth labels.
Unsupervised training enables models to learn to separate the data into distinct classes without explicitly learning what these classes represent.
This method is particularly useful for cases where it is difficult to ascertain the number of classes beforehand. 
Examples of models based on unsupervised learning can be found in \cite{Chan2020, Kalinin2020, Wang2021}. \color{black}


%

Table \ref{tab:models} lists the classification models and image processing algorithms employed by the 20 papers reviewed in this study.


\label{model_selection}

\begin{table}[]
    \centering
    \caption{The class of model and their corresponding training process implemented in a subset of studies reviewed in this work.  }
    \label{tab:models}
    \begin{tabular}{|l|P{7em}|P{12em}| P{5em}|}\hline
        \diagbox[width=10em]{Training\\type}{Model used}& SVM & Neural network & Others \\ \hline
         Training from scratch &\cite{Chowdhury2016,DeCost2015,DeCost2017,Gobert2018,Scime2019}& \cite{Aguiar2019,Ziatdinov2017,Zhu2017,Zhang2021,Yao2020,Yang2019,Cang2016,Baskaran2020,Chen2020,Ciresan2012,Evsevleev2020,Forster2020,Furat2019,Strohmann2019,Stan2020,Roberts2019,Madsen2018,Ma2018,Kaufmann2020,Haan2019} & \cite{Chan2020,Chowdhury2016,Wang2013} \\
         \hline
         Transfer learning & NIL & \cite{Agbozo2019,Azimi2018,Chowdhury2016,Decost2018,Vuola2019,Tsopanidis2020,Hwang2020} & NIL\\
         \hline
    \end{tabular}
\end{table}

\subsection{Model evaluation}

Choosing the appropriate evaluation metric is important in order to obtain a fair assessment of the trained model.
A variety of evaluation metrics can be seen among the published literature in the field.
The choice of a metric is dependent on factors such as the domain-specific problem, size of the dataset, distribution of data among the different candidate classes, etc.
For example, pixel-based classification studies incorporate a distinct set of metrics to evaluate the segmentation performance of different classes.
%
%
It is unfair to compare the performance of two different IDML models with similar domain-specific problem statements, if the models are implemented on different datasets or if they are evaulated using different metrics.  
However, it is useful to understand the different metrics employed by the studies and the rationale behind the choice.
Some of the common evaluation metrics are listed below and are defined based on the following statistics:
\begin{itemize}
  \item True positives, TP$_i$, are data of class label i that were correctly classified as i;
  \item True negatives, TN$_i$, are data not belonging to class label i that were not classified as i;
  \item False positives, FP$_i$, are data not belonging to class label i that were incorrectly classified as i; and
  \item False negatives, FN$_i$, are data of class label i that were not classified as i.
\end{itemize}

\begin{itemize}
     \item For binary classification, average accuracy is defined as
     \begin{equation}
       \frac{TP + TN}{TP + TN + FP + FN}.
     \end{equation}
     This is the most common evaluation metric used, and provides an overall effectiveness of the classifier.  This metric has been used for coarse classification (\cite{Baskaran2020,Chowdhury2016,Aguiar2019,DeCost2017,Forster2020,Gobert2018}) as well as for pixel-level classification (\cite{Strohmann2019,Stan2020,Roberts2019,Chen2020}).  The average accuracy is an appropriate metric when data is uniformly distributed across class labels.
    \item For binary classification, precision (P) and recall (R) are defined as follows:
    \begin{equation}
      P = \frac{TP}{(TP + FP)}
    \end{equation}
    and
    \begin{equation}
      R = \frac{TP}{(TP + FN)}
    \end{equation}
    For multi-class classification, two different ways of averaging are followed: micro-averaging and macro-averaging.  Micro-averaged precision and recall are defined as the application of the respective metric on the cumulative sum of TP, TN, FP, and FN over all the individual classes:
    \begin{equation}
      P_{\mu}= \frac{\sum_{i}TP_i}{\sum_i(TP_i + FP_i)}
    \end{equation}
    Macro-averaged precision and recall are defined as the average of the respective metric applied to each of the individual classes:
    \begin{equation}
      R_M=\frac{\sum_i R_i}{l}
    \end{equation}
    where l is the number of classes.
    Precision and recall are more appropriate metrics than average accuracy when there is an imbalance of data points across different classes.
    For example, in a pixel-level classification to segment defects, the number of defect pixels is likely to be small compared to the number of non-defect pixels or background pixels.
    Examples of studies that have used these two metrics include \cite{Zhu2017,Chen2020,Decost2018,Gobert2018,Vuola2019,Roberts2019}.
    \item The F-score metric, $F_{score}$, derives from the precision and recall values and is defined as follows:
    \begin{equation}
      \frac{(\beta^2+1) P R}{(\beta^2P + R)}
    \end{equation}
    A common value that is used for $\beta$ is 1, and the corresponding F1 score is the geometric mean of the precision and recall.  Examples of studies that have used the F1 metric include \cite{Zhu2017,Chen2020,Tsopanidis2020,Stan2020}.  The F score is important towards in designing a model with a specific trade-off between precision and recall.
    \item Intersection over Union (IoU): This metric is used in applications of object or phase segmentation.  When segmentation is performed through pixel-level classification, IoU can be defined as follows:
    \begin{equation}
      IoU =  \frac{TP}{(TP + FP + FN)}.
    \end{equation}
    Examples of studies that have used this metric include \cite{Azimi2018,Yao2020,Chen2020,Decost2018,Tsopanidis2020,Stan2020,Roberts2019}.
    \item Other metrics:  Some of the studies reviewed in this work have employed evaluation metrics that are unique to the specific problem at hand.
    While the use of such metrics make it difficult to compare the model's performance with similar work, they are a useful measure of the effectiveness of IDML in solving the material science problem.
    Examples of such metrics include difference in predicted and actual interphase connectivity \cite{Strohmann2019}, relative errors in the predicted center of mass and volume of grains as compared to the ground truth values \cite{Furat2019}, comparison of synthesized and original 3-dimensional structures based on spatial correlation metrics such as two-point correlation function and cluster function \cite{Zhang2021}, etc.

\end{itemize}

Table \ref{tab:evaluation} lists the evaluation criteria used by each of the 20 papers reviewed in this study.


\begin{table}[h]
    \centering
    \caption{Evaluation methods adopted by a subset of studies reviewed in this work.  }
    \label{tab:evaluation}
    \begin{tabular}{|P{15em}|P{15em}|}
    \hline
    Evaluation metric  & References  \\
    \hline
    Average accuracy & \cite{Aguiar2019,Ziatdinov2017,Baskaran2020,Chen2020,Chowdhury2016,DeCost2015,DeCost2017,Forster2020,Gobert2018,Strohmann2019,Stan2020,Scime2019,Roberts2019,Ma2018,Kaufmann2020}\\
    \hline
    Precision and Recall & \cite{Zhu2017,Chen2020,Decost2018,Gobert2018,Vuola2019,Roberts2019,Madsen2018} \\
    \hline
    F1 &  \cite{Zhu2017,Chen2020,Tsopanidis2020,Stan2020} \\
    \hline
    IoU & \cite{Azimi2018,Yao2020,Chen2020,Decost2018,Tsopanidis2020,Stan2020,Roberts2019,Hwang2020}\\
    \hline
    Material specific metrics & \cite{Agbozo2019,Zhang2021,Yang2019,Cang2016,Chan2020,Campbell2018,Ciresan2012,Wang2013,Furat2019,Hwang2020,Haan2019} \\
    \hline
    \end{tabular}
\end{table}

\subsection{Integration with the existing workflow}

The final step in leveraging the capability of artificial intelligence to improve materials characterization is integrating the analysis with existing instrumentation or simulation workflows, and thereby handling the implementation challenges.
Most of the studies reviewed in the work have focused on implementing state of the art machine learning or computer vision techniques to a representative dataset of images that exemplify their domain of study.
In contrast, examples of integration of IDML with existing workflows can be seen in the works of Gobert et al. \cite{Gobert2018} and Scime et al.\cite{Scime2019}.
Both these studies have incorporated an imaging source to their additive manufacturing infrastructure in order to perform in-situ analysis of melt pools.
In general, the integration of such techniques with existing workflows present a different set of challenges as compared to the process of adoption of IDML to material datasets.
For example, the domain of electron microscopy is fast emerging as multi-modal and data-intensive \cite{Spurgeon2020}.
The use of high-performance computing infrastructure could enable the speed of analysis to keep pace with the data generation, and also augment the efforts to automate data acquisition through real-time control of instrumentation parameters.
As an example, Patton et al.\cite{Patton2018} demonstrate the implementation of a deep learning network on a GPU supercomputer to extract structural information from atomic-resolution microscopy data.
A software pipeline,built over a supercomputing architecture, to perform segmentation of three-dimensional electron microscopy data directly from the source is discussed in Vescovi et al.\cite{Vescovi2020}.
Seal et al.\cite{Seal2020} report the work in progress towards the use of high-performance computing for segmenting large-scale images.
Factors such as an increasing access to powerful computing resources as well as a focus on developing models that can be run at the edge, i.e., connected to existing instrumentation capabilities, are likely to accelerate the integration. 

\section{Challenges with IDML in the Materials Science Domain}
\label{IDML_challenges}

Although there are several advantages of using an IDML approach in many materials science studies, there remain several challenges that are crucial to the growth and more widespread application of IDML. Among them, small or imbalanced data sets, noisy data, or standard validation techniques are prominent. In addition, there are very limited benchmark data sets widely available for the purpose of comparing different models.

In part, the rapid growth in the machine learning in materials science field has contributed to these challenges, and made it difficult to understand/identify appropriate methods for domain-specific goals, best practices, and opportunities \cite{DaneReview_2020}. To address some of these areas, emerging techniques, best practices, and potential new paths are subsequently discussed.




\section{Emerging techniques and potential new paths}
\label{emerging_practices}
\color{black}

Microstructure characterization using methods from ML and AI has been a nascent endeavour as indicated in the previous sections. The revolution of deep learning to solve seemingly complex computer vision tasks like image recognition has fueled researchers in material science to transfer some of these successes to the field. Section \ref{IDML_current_work} has represented a broad overview of this work from the perspective of application of state of the art machine learning technologies to materials characterization. A lot of this work involves using image datasets of microstructures and other materials characteristics to solve controlled classification problems. Even though this represents a major advancement in the field, several problems still remain as discussed in Section \ref{IDML_challenges}. These challenges can be addressed by emerging methodologies in AI that have the potential to alleviate some of the concerns raised in this study.

One of the challenges in materials characterization is the apparent dearth of labelled datasets. Deep learning problems require massive amounts of labelled data to produce results. Unfortunately, annotating data is time consuming and expensive.  Recent advancements in AI promise methods to tackle this problem such as active learning.  Active learning is inspired from the idea that a predictor that is trained on a smaller set of well-chosen samples can perform as efficiently as a predictor that is trained on a much larger number of randomly sampled data points. In particular, active learning can be implemented as a form of expert-machine interaction. Starting from a small non-optimally sampled batch of data, the algorithm presents the users the images or pixels whose inclusion in the set improves the performance of the predictor. The user or domain expert interacts with the machine in the process of annotation. The procedure is iterated until a stopping criteria has been achieved. Approaches for active learning have been demonstrated in image classification \cite{joshi2009multi}, deep learning \cite{wang2016cost}, remote sensing \cite{tuia2009active}, medical imaging \cite{chowdhury2017active} and image retrieval \cite{tong2001support}. A similar approach can be taken for materials characterization where active learning can be used to reduce annotation burden of material scientists.

A alternate set of methods to reducing the problem of small datasets is known as semi-supervised learning. It considers the problem of supervised learning when only a small subset of data is available for training, which is exactly the challenge that we are looking to overcome. Semi supervised learning algorithms address this problem by learning from unlabelled data as well as labelled data to build better predictors \cite{zhu2005semi}. Semi supervised approaches have been explored for multi-model image classification \cite{guillaumin2010multimodal} and for deep generative modelling \cite{kingma2014semi}. Material characterization will improve with the help of semi-supervised learning algorithms that learn from unlabelled data which is much easier to curate.

A common problem with modern ML and DL methods is that they are inherently inscrutable. Even though they are responsible for reproducing results with astonishing accuracy, they fundamentally lack the ability to provide explanations for their predictions. As a consequence, it is very difficult to interpret the results of a classification or segmentation model. A number of approaches have been proposed recently to tackle this problem. Common techniques for ML model explainability is provided here \cite{gilpin2018explaining}. 
Another way is to visualize the model \cite{zhang2018interpretable, zhang2018visual} by observing the values of filters in trained convolutional neural networks.
Examples of studies which have used these techniques to explain model predictions on material characterization datasets include \cite{Yeung2020} and \cite{Pokuri2019}.
In \cite{Yeung2020}, a CNN was used for prediction of electromagnetic response of nanophotonic structures, following which relevancy-based heatmaps were used to highlight the spatial regions in the input that were important for the prediction. 
Similarly, the technique of computing saliency maps was utilized in \cite{Pokuri2019} to gain insight into the property prediction (short-circuit current) from morphology maps of organic photovoltaic films. 
Using such techniques to understand the latent representation of data and the trained state of hidden layers could accelerate the process of material design and discovery. \color{black}
Another viewpoint of interpretability is by reducing model complexity. The core idea is that simpler models are easier to understand than larger esoteric models. For example,  \cite{ribeiro2016should} tries to approximate the complex model with a locally linear model. A related idea is using feature importance \cite{lundberg2017unified} for interpreting predictions. Symbolic approaches \cite{lazaridou2016multi, chowdhury2020escell, santamaria2020towards} generate symbolic expressions to generate semantic understanding of the predictions. 
An example of an interpretable workflow for material characterization is \cite{Liu2020}, in which a regression model is developed to predict peak stress from a morphology image generated from SEM. Using a generative method, structural features (such as porosity) are systematically modified to study its effect on the stress. This analysis enables one to identify the morphological features that are key for a targeted output. \color{black}
As discussed before, these challenges involving interpretability need to be addressed before materials characterization using IDML is to be widely accepted.

\color{black}

\subsection{Generative methods}


Before GANs, there have been many generative models, such as the Deep Boltzmann Machine (DBM) and Variational Auto-encoder (VAE). However, they had less of an impact due to the difficulty of of approximating intractable probabilistic computations in maximum likelihood estimation, known as the ``maximum likelihood training paradigm’’, and leveraging the benefits of piecewise linear units in the generative context \cite{NIPS2014_5423}. Unlike previous generative models, Generative Adversarial Network (GAN) brings a new breath to the deep generative models with its simple structure and effectiveness in image generation. With the help of Convolutional Neural Networks, the GANs can generate realistic images, as large as 1024 by 1024 pixels.

A GAN framework consists of a generator, $G$, that generates samples from a noise variable, $z$, and a discriminator,$D$, that aims to distinguish between samples from the real data distribution and those from the synthetic data distribution (from the generator). The generator can be thought of as counterfeiters trying to produce fake but realistic images, while the discriminator aims to distinguish the fake images from the real ones.

GANs have been proven successful for many image synthesis and unsupervised learning tasks. It is a popular framework for representation learning, such as disentangle pose from lighting in 3D rendered images \cite{NIPS2016_6399}, and image completion, where large missing regions are synthesized utilizing the surrounding image features \cite{8578675}. Variants of GANs have surpassed many other generative models in the quality of samples as well as their underlying representation. GANs have also been applied to the task of image-to-image translation, which learns a mapping between an input image and an output image. Several examples of this image-to-image translation exist in literature, including \cite{8237506} used CycleGAN to transfer images of one style to another, such as between landscape images in summer and in winter, and between photographs and paintings of Monet. \cite{8953676} used GauGAN to create photorealistic images from segmentation maps, which are labeled sketches that depict the layout of a scene. \cite{DBLP:journals/corr/IsolaZZE16} used Pix2Pix on many applications of image-to-image translation, such as mapping from aerial photos to maps, and mapping from edges to photos \cite{DBLP:journals/corr/IsolaZZE16}.

\begin{figure}
    \centering
	\includegraphics[width=0.75\textwidth]{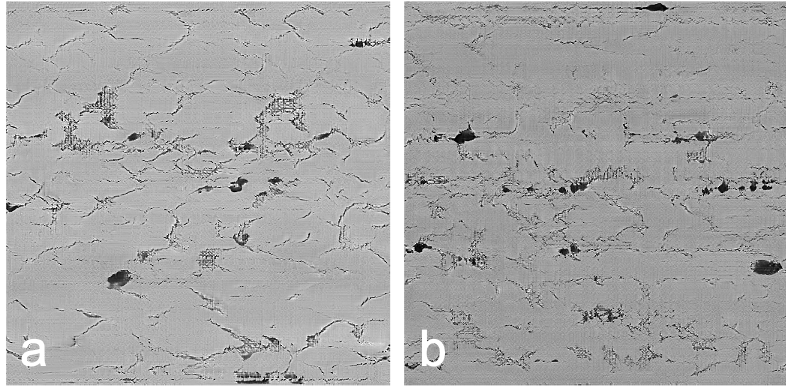}
	\includegraphics[width=0.75\linewidth]{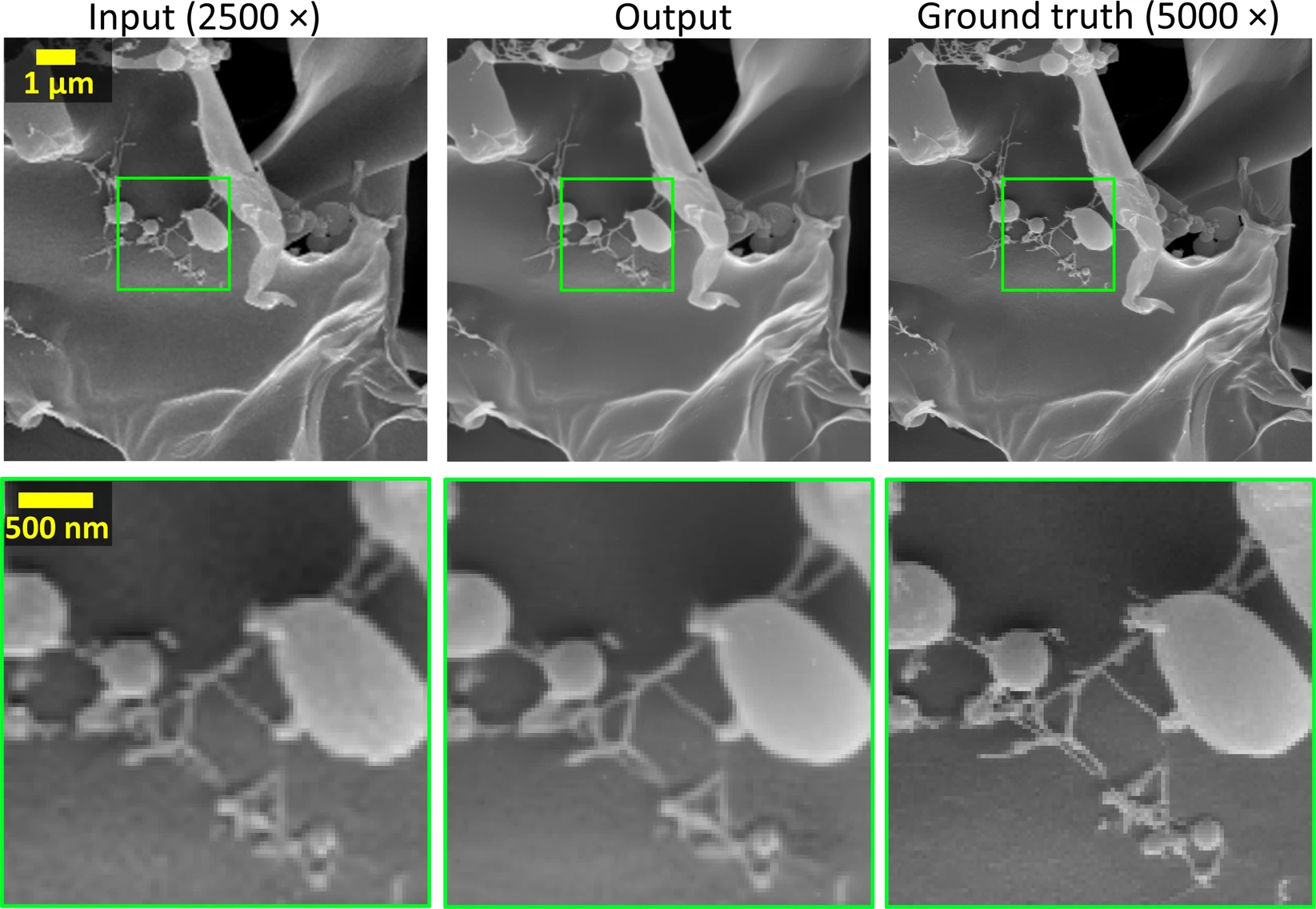}
	\caption{Two types of applications in the field of materials characterization that have benefited from using generative models. (Top) Example synthetic images generated by the trained Progressive Growing GAN. The generated images show varying microstructural features, specifically different extent of lamellar transformation products, and distribution of carbides. Figure reproduced with permission from \cite{Ma2020_1}; Creative Common CC BY. (Bottom) The use of a generative model to enhance the resolution of an image acquired from a Scanning Transmission Electron Microscope. The benefit of this application lies in a reduced imaging time as well as a reduced likelihood for electron beam damage to the samples. Figure reproduced with permission from \cite{Haan2019} ; Creative Commons Attribution 4.0 International. } \label{fig:pggan-samples}
\end{figure}

Recently, GANs have emerged as a promising methodology for application in computational materials design, for the purpose of developing structure-property and structure-performance relations via physical simulations \cite{Yang2018GAN, Li2018_adversarial, Ma2020_1}. With the help of CNN backbone, GANs are able to capture complex microstructural characteristics. In \cite{Mosser2017}, GAN is used to generate high-resolution three-dimensional images of the pore space at different scales. In \cite{Yang2018GAN}, the authors proposed a deep adversarial learning methodology to overcome the limitations of existing microstructure characterization and reconstruction (MCR) techniques. In this work, the GAN is used identify the mapping between latent codes and microstructures. In \cite{Li2018_adversarial}, the GANs are used to learn a mapping between latent variables and microstructures, which is later used as the design variables to obtain microstructures of desired material property with the help of a Bayesian optimization framework. In \cite{Ma2020_1}, the authors generated high-resolution microstructure images with the advanced progressive-growing GAN (pgGAN), and interpret the task of microstructure generation as microstructure distribution characterization and image-to-image translation.
Figure \ref{fig:pggan-samples} highlights the two types of applications that have benefited from using generative models. 

\subsection{Merging data from multiple characterization techniques}

\begin{figure}[H]
    \centering
    \includegraphics[width = \textwidth, height = 1.25in]{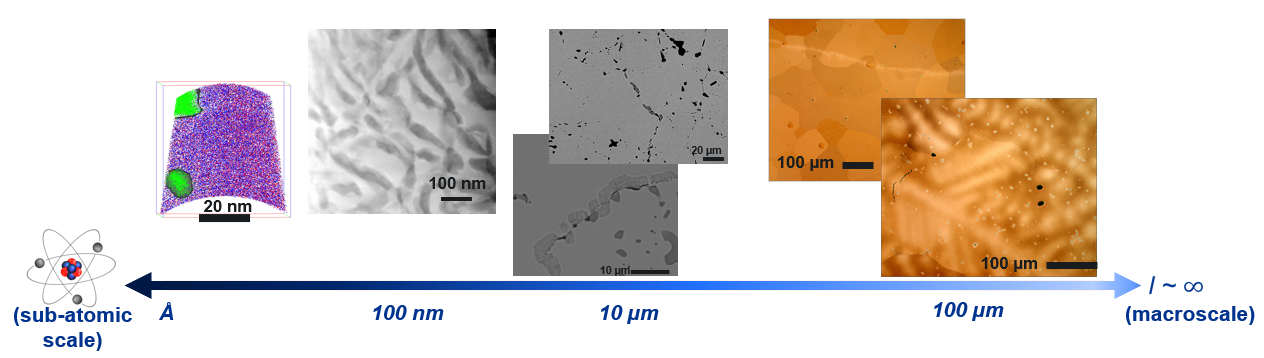}
    \caption{Schematic of data from different image types that span across length scales and are instrumental to materials characterization studies. From left to right: atom probe tomography data, STEM bright field image, SEM micrographs, optical micrographs.}
    \label{fig:lengthscales_schematic}
\end{figure}

Materials characterization studies typically involve the application of multiple experimental techniques, sometimes in combination with modelling or computation, in order to develop microstructure-processing-property relationships. The multi-length scale nature of such studies necessitates varying data types to characterize any microstructure, which is schematically shown in Figure \ref{fig:lengthscales_schematic}. The multiple data types shown here span from atom probe tomography data (sub-nanometer scale spatial resolution, typically analyzing volumes to 50 nm by 50 nm by 100 nm) to TEM images, to SEM and optical micrographs. These data types are only a few examples of the data types used in materials characterization. Considering a more specific example, of analyzing an alloy after casting. This alloy may then be subjected to a homogenization anneal, and subsequently several thermo-mechanical processing to achieve the desired dimensions and properties. To examine the microstructures produced after processing, a researcher may begin with cross-sectioning and polishing the material and examining the cross-section using an optical microscope. Some microstructural features may be visible, depending on the microstructure (i.e. grain boundaries, inclusions, etc.), but others may not, hence the researcher will go to the SEM to look at the same microstructure with higher resolution. Secondary electron (SE) and back scatter electron (BSE) images may be collected, all images, but providing different detail about the microstructure (i.e. topography for SE images versus atomic number contrast for BSE images). Energy dispersive spectroscopy (EDS) may be performed to obtain element distributions or composition information from varying regions of the microstructure. EDS data can be collected in points, or as maps. Now, multiple image and data types have been collected, all providing unique detail about the microstructure in question. Next, electron backscatter diffraction (EBSD) may be collected to provide grain orientation information, and another data type and instance is provided. For even higher resolution characterization, transmission electron microscopy (TEM) may be employed to identify structure, and perform EDS mapping and point analysis in small volumes, with a typical sample size of ~10\textmu m by 10\textmu m with less than 100 nm thickness (to be electron transparent). Higher resolution chemical and isotopic mapping can also be performed using atom probe tomography (APT), yielding a three dimensional point cloud data set with the x,y,z position and elemental and isotopic identity of each atom detected. The workflow detailed here is just one example, with limited techniques. Several other techniques exist and are selected based on the material of interest and characterization challenge faced. The methods discussed here are limited to characterization of a material via microscopy methods that yield data primarily in the form of images, without regard to thermal or mechanical properties.
Yet, the key concept demonstrated through this example is sound, and remains a critical component to the future of machine learning in materials characterization and design: a single material condition can be described by multi-lengthscale, multi-modal data sets. 
{While many IDML research works have been performed using images from a single imaging modality, the concept of fusing multiple imaging modalities is still in its infancy in materials science. Multimodal learning is when learning algorithms relate information from multiple data collection sources, as detailed by   Ngiam et al \cite{Ngiam2011}, who focused on relating speech audio to videos of lip movement. In the materials science domain, a multimodal learning problem could involve data collected using multiple different chemical imaging modalities in order to relate information such as crystal structure to composition, for example. Independent of application area, multimodal data fusion aims to capture correlations across modalities, such as from SEM images that detail microstructural features of interest to TEM and APT that detail structure and composition of phases present, grain boundary character, etc. Multimodal chemical imaging, correlative and complementary microscopy experiments are uniquely powerful since they allow for correlation of properties with chemical composition and structure, which is central to understanding how these drive material performance \cite{Belianinov2018}.}  

\section{Summary}

In this work, we have reviewed the field of image driven machine learning (IDML) towards the analysis of material structure characterization.
The scope of the discussion included studies based on sample surface images from cameras, digital images from optical and electron microscopes, diffraction pattern images, and simulated images of material structures at various length scales.
A canonical series of six action steps was defined, recognizing the importance of multiple independent stages required for a successful application of IDML in the field: problem definition, dataset building, model selection, model training, model evaluation, and integration with the existing workflow.
%
Two of the most widely adopted characterization problems have been morphology class detection using image classification (one class label per image) and phase segmentation for quantitative analysis using pixel-level classification (one label per pixel).
An emerging sub-field in the last couple of years has been the adoption of generative models to augment the image acquisition process, especially useful for high-resolution and destructive characterization techniques such as electron microscopy.
In the last few years, the field has seen the widespread adoption of neural network models such as the convolutional neural network for image classification as well as for semantic segmentation through pixel-level classification.
Though the difficulty in acquiring high quality data in the field is a challenge to training complex models, the rise in adoption of the training paradigm of transfer learning promises to partially alleviate this limitation.
In large part, the material science community has thus far shown the utility of image driven machine learning as a useful paradigm to characterize material structures.
With increasing access to high-performance computing and data architectures, robust model development through the incorporation of interpretability and explainability, the next stage of progress for the field will entail the integration of IDML into existing instrumentation and simulation workflows.

\section{Acknowledgements}

This work was in part supported by the U.S. Department of Energy (DOE) National Nuclear Security Administration, and in part supported by the National Science Foundation ; Project award No. CMMI-1729336. A portion of this work was performed at the Pacific Northwest National Laboratory (PNNL), which is operated for the U.S. DOE by Battelle Memorial Institute under Contract No. DE-AC05-76RLO1830.

\end{document}